\documentclass[letterpaper]{article}
\usepackage{aaai}
\nocopyright
\usepackage{times}
\usepackage{helvet}
\usepackage{courier}
\usepackage{amsfonts,amssymb}
\usepackage{stmaryrd}
\usepackage{amsmath}

\newcommand{\OObj}{{\cal J}}
\newcommand{\AAtr}{{\cal U}}
\newcommand{\Obj}{{\sf J}}
\newcommand{\Atr}{{\sf U}}
\newcommand{\Rat}{{\sf R}}
\newcommand{\Tr}{{\rm tr}}
\newcommand{\Prob}{{\rm P}}
\newcommand{\Sim}{{\rm s}}

\newcommand{\Exp}{{\bf E}}
\newcommand{\bal}[1]{{#1}^0}
\newcommand{\nor}[1]{\underline{#1}}

\newcommand{\Ww}{M}

\newcommand{\AAA}{{\cal A}}
\newcommand{\BBB}{{\cal B}}

\newcommand{\LLL}{{\cal L}}

\newcommand{\PPP}{{\cal P}}

\renewcommand{\Bbb}{\mathbb}

\newcommand{\BBb}{{\Bbb B}}
\newcommand{\CCc}{{\Bbb C}}

\newcommand{\NNn}{{\Bbb N}}

\newcommand{\RRr}{{\Bbb R}}

\newcommand{\ZZz}{{\Bbb Z}}

\newcommand{\WP}{\PPP}

\newcounter{countroman}
{\begin{list}{{\rm (\roman{countroman})}}{\usecounter{countroman}}}%
{\end{list}}

\newcounter{countalpha}
{\begin{list}{(\alph{countalpha})}{\usecounter{countalpha}}}%
{\end{list}}

\newcounter{countalphabf}
{\protect\begin{list}{{\rm (}{\bf \protect\alph{countalphabf}}{\rm%
)}}{\protect\usecounter{countalphabf}}}%
{\end{list}}

\mathcode`\<="4268 
\mathcode`\>="5269 
\mathchardef\gt="313E 
\mathchardef\lt="313C 

 \def\pushright#1{{
    \parfillskip=0pt            
    \widowpenalty=10000         
    \displaywidowpenalty=10000  
    \finalhyphendemerits=0      
    \leavevmode                 
    \unskip                     
    \nobreak                    
    \hfil                       
    \penalty50                  
    \hskip.2em                  
    \null                       
    \hfill                      
    {#1}                        
    \par}}                      

 \def\qed{\pushright{$\square$}\penalty-700 \smallskip}

\newenvironment{prf}[1]{\begin{trivlist} \item[{\bf ~Proof}#1.]}%
{\qed\end{trivlist}}

\newcommand{\beq}{\begin{equation}}
\newcommand{\eeq}{\end{equation}}
\newcommand{\ba}[1]{\begin{array}{#1}}
\newcommand{\ea}{\end{array}}
\newcommand{\bea}{\begin{eqnarray}}
\newcommand{\eea}{\end{eqnarray}}
\newcommand{\bear}{\begin{eqnarray*}}
\newcommand{\eear}{\end{eqnarray*}}
\newcommand{\bpr}{\begin{prf}{}}
\newcommand{\epr}{\end{prf}}
\newcommand{\bprf}[1]{\begin{prf}{#1}}
\newcommand{\eprf}{\end{prf}}

\newcommand{\ot}{\xymatrix@C-.5pc{& \ar[l]}}
\newcommand{\tto}[1]{\xymatrix@C-.5pc{\ar[r]^{#1}&}}
\newcommand{\oot}[1]{\xymatrix@C-.5pc{&\ar[l]_{#1}}}
\newcommand{\mono}{\xymatrix@C-.5pc{\ar@{>->}[r]&}} 
\newcommand{\mmono}[1]{\xymatrix@C-.5pc{\ar@{>->}[r]^{#1}&}} 
\newcommand{\eepi}[1]{\xymatrix@C-.5pc{\ar@{->>}[r]^{#1}&}}
\renewcommand{\mapsto}{\xymatrix@C-.5pc{\ar@{|->}[r]&}}
\newcommand{\mmapsto}[1]{\xymatrix@C-.5pc{\ar@{|->}[r]^{#1}&}}
\newcommand{\inclusion}{\xymatrix@C-.5pc{\ar@{^{(}->}[r] &}}
\newcommand{\iinclusion}[1]{\xymatrix@C-.5pc{\ar@{^{(}->}[r]^{#1}&}}
\newcommand{\dtto}[2]{\xymatrix@C-.5pc{\ar@<.5ex>[r]^{#1} \ar@<-.5ex>[r]_{#2}&}}

\newcommand{\id}{{\rm id}}
\newcommand{\pfn}[3]{\xymatrix@-.8pc{{#1}\ar@{->}[r]|-{\circ}^-{#2}&{#3}}}
\newcommand{\Pfn}[2]{\xymatrix@-1pc{{#1}\ar@{=>}[r]|-{\circ}&{#2}}}
\newcommand{\fn}[3]{\xymatrix@-.8pc{{#1}\ar@{->}[r]^-{#2}&{#3}}}

\newcommand{\rel}[3]{\xymatrix@-.8pc{{#1}\ar@{->}[r]|-{\scriptscriptstyle |}^-{#2}&{#3}}}
\newcommand{\rell}[3]{\xymatrix{{#1}\ar@{->}[r]|-{\scriptscriptstyle \circ}^-{#2}&{#3}}}

\usepackage{xy} 
\xyoption{all} 
\CompileMatrices 

\title{On quantum statistics in data analysis 
}
\author{Dusko Pavlovic\thanks{Supported by EPSRC and ONR.}\\ Kestrel Institute and Oxford University\\{\tt dusko@\{kestrel.edu,comlab.ox.ac.uk\}}
}
\date{}

\begin{document} 
\maketitle

\begin{abstract} 
Originally, quantum probability theory was developed to analyze statistical phenomena in quantum systems, where classical probability theory does not apply, because the lattice of measurable sets is not necessarily distributive. On the other hand, it is well known that the lattices of concepts, that arise in data analysis, are in general also non-distributive, albeit for completely different reasons. In his recent book, van Rijsbergen \shortcite{vanRijsbergenGIR} argues that many of the logical tools developed for quantum systems are also suitable for applications in information retrieval. I explore the mathematical support for this idea on an abstract vector space model, covering several forms of data analysis (information retrieval, data mining, collaborative filtering, formal concept analysis\ldots), and roughly based on an idea from categorical quantum mechanics \cite{Abramsky-Coecke,PavlovicD:QMWS}. It turns out that quantum (i.e., noncommutative) probability distributions arise already in this rudimentary mathematical framework. We show that a Bell-type inequality  \cite{Bell64EPR} must be satisfied by the standard similarity measures, if they are used for preference predictions. The fact that already a very general, abstract version of the vector space model yields simple counterexamples for such inequalities seems to be an indicator of a genuine need for quantum statistics in data analysis.
\end{abstract}

\section{Introduction}
Until recently, Computer Science was mainly concerned with
data storage and processing in purpose-built data bases and computers. With the advent of the Web and social computation, the task of finding and understanding information arising from local interactions in spontaneously evolving computational networks and data repositories has taken center stage.

As computers evolved from calculators, the key paradigm of Computer
Science was computation-as-calculation, with the Turing Machine construed as a generic calculator, and with data processing performed by a small set of local operations. As computers got connected into networks, and captured a range of social functions, the paradigm of computation-as-communication emerged, with data processing performed not only locally, but also through distribution, merging, and association of data sets through various communicating processes. Such non-local data processing has been implemented through markets, elections, and
many other social mechanisms for a very long time, albeit on a smaller scale, with less concrete infrastructure, and with more complex computational agents. A new family of its implementations is based on a new computational platform, which is not any more the Computer, or even its operating system, but the Web, and its knowledge systems.

But while the interfaces of the local computational processes are defined to be the interfaces of the computers which perform them, the carriers of computation-as-communication do not come with clearly defined interfaces. The task of finding and supplying reliable data within a market, or on the Web, or in a social group, carries with it many deep problems. Two of them are particularly relevant for this work.

\subsection{Problem of partial information and indeterminacy}
Data processing in a network is ongoing. On the other hand, the data sets are usually incomplete, and information needs to be extracted from such incomplete sets. E.g., a task in a recommender system is to extrapolate which movies (books, music\ldots) will a user like, from a sparse sample of those that she had previously rated. In information retrieval, the task is to extrapolate which information is relevant for a query, from a small set of tokens characterizing the query on one hand, and the information on the other hand.

In the standard model of data analysis, succinctly presented e.g. in \cite{azar01spectral}, it is assumed that a matrix of random variables, containing a complete information about the relevant properties of the objects of interest, exists out there (in some sort of a Platonic heaven of information), and can be sampled. The problem of data analysis is that the sampling process is noisy, and partial; more specifically, that the distributions of the random variables are distorted by an error process, and by an omission process. The task of data analysis is to eliminate the effects of these processes, and reconstruct a good approximation of the original information. 

While mathematically convenient, and computationally effective, this model does not seem very realistic. If we instantiate it to a recommender system again, then its basic assumption becomes that each user has a completely defined preference distribution, albeit only over the items that he has used, and that the recommender system just needs to reconstruct this preference distribution. But if we zoom in, and ask the user himself, he will often be unable to precisely reconstruct his own preference distribution. If we ask him to rate some items again, he will often assign different ratings. One reason is that information processing is ongoing, and that the preferences evolve and change. If we zoom in even further, we will find that the state of user's preferences is usually not completely determined even in a completely static model: right after watching a movie, one usually needs to toss a "mental coin" to decide whether to assign 2 or 3 stars, say, to the performance of an actor; or to decide whether to pay more attention, while watching the movie, to this or that aspect, music, colors\ldots 

While the indeterminacy of information in a network can be reduced to an effect of noise, like in the standard model, and averaged out, it is interesting to ponder whether viewing this indeterminacy as an essential feature of network computation, rather than a bug, may lead to more realistic models of information systems. Is the "mental coin", which resolves the superposition of the many components of my preferences when I need to measure them, akin to a real coin, which we all agree is governed by completely deterministic laws of classical physics, and its randomness is just the appearance of its complex behavior; or is this "mental coin" governed by a more fundamental form of randomness, like the one that occurs in quantum mechanics, causing the superposition of many states  to {\em collapse under measurement}? 

\subsection{Problem of classification and latent semantics} 
The task of conceptualizing data has been formulated in many ways. In information retrieval, the central task is to determine the relevance of data with respect to a query. In recommender systems, the implicit query is always: "What will I like, given my past choices and rankings?", and the task is to find the relevant recommendations. In order to tackle such tasks, one classifies the data on one hand, the queries on the other, and aligns the two classifications, in order to extrapolate the future choices from the past choices. --- But what are these classifications based on? 

The simplest approach is based on keywords. But even classifying a corpus of purely textual documents, viewed as bags of words, according to the frequency of the occurrences of the relevant keywords, leads to significant problems: polysemy, homonymy, synonymy.  The problem becomes very difficult when it comes to classifying families of non-textual objects: images, music, video, film. Only a small part of their correlations can be captured by connecting the keywords, captions, or other forms of textual annotations.

Latent semantics correlates data by extracting their intrinsic structure. For instance, the central piece of the original Google search engine, distinguishing it from other similar engines, was that the keyword search was supported by PageRank  \cite{page98pagerank}, a reputation ranking of the Web pages, extracted from their intrinsic hyperlink structure. Even for the keyword search, the crucial step was to recognize this latent variable \cite{Everitt84latent} extracting relevance from non-local network structure, rather than from local term occurrence. Such semantical support is even more critical for search and retrieval of non-textual information, on the Web and in other data spaces. 

\section{Overview of latent semantics}
We consider the case when two types of data assign the meaning to each other.

\subsection{Pattern matrices}
Latent semantics is generally given as a map
\bear
\Obj \times \Atr &\tto{A} & \Rat
\eear
where 
\begin{itemize}
\item $\Obj$ is a set of {\em objects}, or {\em items},
\item $\Atr$ is a set of {\em properties}, or {\em users},
\item $\Rat$ is a set of {\em values}, or {\em ratings}.
\end{itemize}
This map is conveniently presented as a {\em pattern matrix\/} $A = (A_{iu})_{\Obj \times \Atr}$. The entry $A_{iu}$ can be intuitively written as a model relation $i\models u$, especially when $\Rat = \{0,1\}$. In general, it can be construed as the degree to which the object $i$ satisfies the property, or the user $u$. While the ratings $\Rat$ usually carry a structure of an ordered {\em rig}\footnote{A {\em rig\/} $\Rat = (\Rat,+,\cdot,0,1)$ is a "ring without the negatives". This means that $(\Rat,+,0)$ and $(\Rat,\cdot,1)$ are commutative monoids satisfying $a(b+c) = ab+ac$ and $0a=0$. The typical examples include natural numbers, non-negative reals, but also distributive lattices, which generally do not embed in a ring.}, the attributes $\Atr$ often carry a more general algebraic structure, whereas the behaviors of the objects in $\Obj$ may be expressed coalgebraically. Clearly, the rig structure of $\Rat$ is just enough to support the usual matrix composition. Sometimes, but not always, we also assume that $\Rat$ has no nilpotents, so that it can be embedded in an ordered field.

\subsubsection{Examples.} \hspace{2em}
\vspace{.5\baselineskip}

{\tiny
\begin{tabular}{|c||c|c|c|c|}
\hline
domain & $\Obj$ & $\Atr$ & $\Rat$ & $A_{iu}$\\
\hline \hline 
text analysis & documents & terms & $\NNn$ & occurrence count \\
\hline
measurement & instances & quantities & $\RRr$ & outcome \\
\hline
user preference & items & users & \{0,\ldots,5\} & rating \\
 \hline
 topic search & authorities & hubs & $\NNn$ & hyperlinks \\
 \hline
concept analysys & objects & attributes & \{0,1\} & satisfaction \\
\hline
elections & candidates & voters & \{0,\ldots,n\} & preference \\
\hline
market & producers & consumers  & $\ZZz$ & delivery \\
\hline
digital images & images & pixels & $[0,1]$ & intensity \\
\hline
\end{tabular}
}
 
 \subsection{Balancing and normalization}
\subsubsection{Notation.} For every vector $x = (x_k)_{k=1}^n$, we define \begin{itemize}
\item the average (expectation) $\Exp(x)  =  \frac{1}{n}\sum_{k=1}^n x_k$
\item the $\ell_2$-norm $\lVert x\rVert_2 = \sqrt{\sum_{k=1}^n |x_k|^2}$,
\item the $\ell_\infty$-norm $\lVert  x\rVert_\infty = \bigvee_{k=1}^n |x_k|$.
\end{itemize}

\subsubsection{Item balancing} \hspace{-1em} of a semantics matrix $A$ reduces each of its rows $A_{i\bullet}$, corresponding to the item $i$, to a row vector $\bal{A}_{i\bullet}$, defined
\bear
\bal{A}_{i\bullet} & = & A_{i\bullet} - \Exp(A_{i\bullet})
\eear 
The unassigned ratings in $A_{i\bullet}$ are padded by zeros.

In an item-balanced matrix records, the difference between the items with a higher average rating and the items with a lower average rating is factored out. Only the satisfaction profile of each item is recorded, over the set of users who have assigned it better-than-average, or worse-than-average rating. The average and unassigned ratings are identified, and both become 0. 

\subsubsection{User balancing} \hspace{-1em} of a semantics matrix $A$ reduces each of its columns $A_{\bullet u}$, corresponding to the user $u$, to a column vector $\bal{A}_{\bullet u}$, with the expected value 0, by setting
\bear
\bal{A}_{\bullet u} & = & A_{\bullet u} - \Exp(A_{\bullet u})
\eear 
The unassigned ratings are again padded by zeros.

In a user-balanced matrix, users' different rating habits, that some of them are more generous than others, are factored out. Only the satisfaction profile of each user is recorded, over the set of all items that she has rated. The average and unassigned ratings are identified, both with 0. 

\subsubsection{Item normalization} \hspace{-1em} of a semantics matrix $A$ factors its rows into unit vectors; the {\bf user normalization} factors its columns into unit vectors --- by setting
\bear
\nor{A}_{i\bullet} & = & \frac{A_{i\bullet}}{\lVert A_{i\bullet}\rVert_2}\\
\nor{A}_{\bullet u} & = & \frac{A_{\bullet u}}{\lVert A_{\bullet u}\rVert_2}
\eear

\subsubsection{Comment.} The purpose of balancing and normalization of raw semantic matrices is to factor out the aspects of rating that are irrelevant for the intended analysis. Whether a particular adjustment is appropriate or not depends on the intent, and on the available data. E.g., padding the available ratings by assigning the average rating to all unrated items may be useful in some cases, but it skews the data when the sample is small.\footnote{E.g., when only one rating is available from a user, then extrapolating his average rating to the unrated items simply erases all available information.}  {\em In the rest of the paper, we assume that all such adjustments have been applied to data as appropriate, and we focus on the methods for extracting information from them.}

\subsection{Classification}
Through pattern matrices and latent semantics, the objects and the properties lend a meaning to each other. The simple method for extracting that meaning is based on the general ideas of Principal Component Analysis \cite{Jolliffe86PCA}. This method underlies not only the vector space based approaches, like Latent Semantics Indexing (LSI) \cite{Deerwester90indexing}, or Hypertext Induced Topic Search (HITS) \cite{kleinberg99authoritative}, but also, albeit in a less obvious way, Formal Concept Analysis (FCA) \cite{Wille82FCA}, and some other approaches. The general idea is that the latent semantical structures can be obtained by factoring the pattern matrix through suitable transformations, required to preserve a {\em conceptual distance\/} between the objects, as well as between their properties. These distance-preserving transformations can be captured under the abstract notion of {\em isometry}.

Suppose that the rig of values is given with an involutive automorphism $\overline{(-)}:\Rat\to\Rat$, called {\em conjugation}. If the values are the complex numbers, $\Rat = \CCc$, then of course $\overline{a+ib} = a-ib$. For general rigs $\Rat$, conjugation sometimes boils down to $\overline{a} = a$. In any case, any pattern matrix $A = (A_{iu})_{\Obj\times\Atr}$ induces an {\em adjoint\/} matrix $A^\ddag = (A^\ddag_{ui})_{\Atr\times \Obj}$, whose entries are defined to be $A^\ddag_{ui} = \overline{A}_{iu}$. The {\em inner product\/} of vectors $x,y\in \Rat^\Obj$ can now be defined as $<x|y> = y^\ddag \circ x$.

\paragraph{Definitions.} An {\em isometry\/} is a map $U:\AAA\inclusion \BBB$ such that $<Ux|Uy> =<x|y>$ holds for all $x,y$. Equivalently, this means that $U^\ddag U = \id_\AAA$. It is a  {\em unitary\/} if both $U$ and $U^\ddag$ are isometries.

An {\em isometric decomposition\/} of an operator $B:\AAtr\to \OObj$ consists of isometries $\hat{V}:\hat{\AAtr} \inclusion \AAtr$ and $\hat{W}:\hat{\OObj}\inclusion\OObj$ such that there is a (necessarily unique) map $\hat{B}:\hat{\OObj}\to \hat{\AAtr}$ satisfying $B  =  \hat{W} \hat{B} \hat{V}^\ddag$
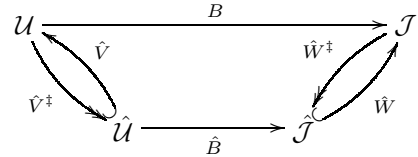
\[\xymatrix{
\AAtr \ar[rrrr]^B \ar@/_/
@{->>}[dr]_{\hat{V}^\ddag}  && && \OObj \ar@/_/
@{->>}[dl]_{\hat{W}^\ddag} \\
 & \hat{\AAtr} \ar[rr]_{\hat{B}} \ar@/_/
 @{^{(}->}[ul]_{\hat{V}} & & \hat{\OObj} \ar@/_/
 @{^{(}->}[ur]_{\hat{W}}
}\]
The {\em spectral decomposition\/} $B = \bar{W} \bar{B} \bar{V}^\ddag$ is minimal among $B$'s isometric decompositions:
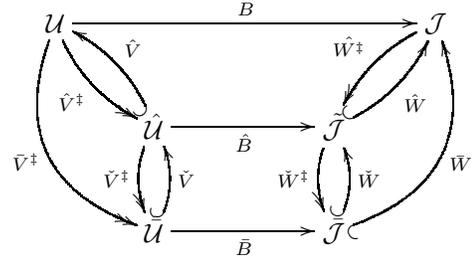
\[\xymatrix{
\AAtr \ar[rrrr]^B \ar@/_2pc/@{->>}[ddr]_{\bar{V}^\ddag}
\ar@/_/@{->>}[dr]_{\hat{V}^\ddag}  && && \OObj \ar@/_/@{->>}[dl]_{\hat{W^\ddag}} \\
 & \hat{\AAtr} \ar[rr]_{\hat{B}} \ar@/_/@{^{(}->}[ul]_{\hat{V}} 
 \ar@/_/@{->>}[d]_{\check{V}^\ddag} 
 & & \tilde{\OObj} \ar@/_/@{->>}[d]_{\check{W}^\ddag}
 \ar@/_/@{^{(}->}[ur]_{\hat{W}}\\
 &  \bar{\AAtr} \ar@/_/@{^{(}->}[u]_{\check{V}}  \ar[rr]_{\bar{B}} && \bar{\OObj} \ar@/_/@{^{(}->}[u]_{\check{W}} \ar@/_2pc/@{_{(}->}[uur]_{\bar{W}}
}\]
in the sense that for every isometric decomposition $B = \hat{W} \hat{B} \hat{V}^\ddag$, 
there is an isometric decomposition $\hat{B} = \check{W} \bar{B} \check{V}^\ddag$, such that $\bar{W} =\hat{W}\check{W}$ and $\bar{V} = \hat{V}\check{V}$.

We further also need

\subsubsection{Correlation matrices }\hspace{-1.5ex} are the self-adjoint matrices in the form $\Ww^\Obj = AA^\ddag$ and $\Ww^\Atr = A^\ddag A$, i.e.
\bear
\Ww^\Obj_{ij} & = & \sum_{u\in \Atr} \overline{A}_{ju}\cdot A_{iu}\\
\Ww^\Atr_{uv} & = & \sum_{i\in \Obj} A_{iu}\cdot \overline{A}_{iv}
\eear

\subsection{Examples of classification through isometric decomposition}
Given a pattern matrix $\Obj \times \Atr \tto{A} \Rat$, we set
\bear
\OObj & = & \Rat^\Obj\\
\AAtr & = & \Rat^\Atr
\eear
so that $A$ becomes a linear operator $A:\AAtr\to \OObj$, defined by the usual matrix action on the vectors.

\paragraph{Latent Semantic Indexing.} \cite{Deerwester90indexing} Let the rig of values $\Rat$ be the field of real numbers $\RRr$, with the trivial conjugation $\overline{r} = r$. This means that $\OObj = \Rat^\Obj$ and $\AAtr = \Rat^\Atr$ are real vector spaces. The pattern matrix $\Obj \times \Atr \tto{A} \Rat$ induces the linear operator $\AAtr \tto{A} \OObj$ and the adjoint $\OObj \tto{A^\ddag} \AAtr$ is just the transpose.

The isometric decomposition boils down to the singular value decomposition. The isometries $V:\AAtr' \inclusion\AAtr$ and $W:\OObj'\inclusion\OObj$ are obtained by the spectral decomposition of the symmetric matrices $\Ww^\Atr = A^\ddag A$ and $\Ww^\Obj = AA^\ddag$. Since both decompose through the same rank space, with the same spectrum $\Lambda = \{\lambda_1\geq \lambda_2\geq \ldots \geq\lambda_n\}$, we get a positive diagonal matrix $\Lambda$ such that  $ A^\ddag A = V\Lambda V^\ddag$ and $AA^\ddag = W\Lambda W^\ddag$, from which $A = WDV^\ddag$ follows for $D = \sqrt{\Lambda}$. 
%

The eigenspaces of $\Ww^\Atr$ and $\Ww^\Obj$ can be viewed as {\em pure topics\/} captured by the pattern matrix $A$. The eigenvalues correspond to the degree of semantical relevance of each topic in the data set from which the pattern matrix was extracted. If $\Atr$ are users and $\Obj$ items, then the eigenspaces in $\AAtr$ can be thought of as {\em tastes}, the eigenspaces in $\OObj$ as {\em styles}. Remarkably, there is a bijective correspondence between the two, and the eigenvalues quantify the correlations. As an instance of the same decomposition, Kleinberg's \shortcite{kleinberg99authoritative} analysis of Hyperlink Induced Topic Search (HITS) yields a similar correspondence between the hubs and the authorities on the Web. In all cases, the underlying view is that the information consumers and the information producers, lending each other the latent semantics, share a uniform conceptual space. An even simpler presentation of that optimistic view is

\paragraph{Formal Concept Analysis.}  \cite{FCA} Let the rig of values  $\Rat$ now be the distributive lattice $\BBb= (2,\vee,\wedge,0,1)$, over the underlying set $2= \{0,1\}$, with the negation $\neg: \BBb \to \widetilde{\BBb}$ as the conjugation $\overline{\imath} = \neg i$. Note that this is now an antimorphism of $\BBb= (2,\vee,\wedge,0,1)$ with the dual lattice $\widetilde{\BBb}= (\widetilde{2},\wedge,\vee,1,0)$. The space of the objects is thus the boolean lattice $\OObj = 2^\Obj$, ordered by inclusion, whereas the space of the properties is the boolean lattice $\AAtr = \widetilde{2}^\Atr$, ordered by reverse inclusion. 

Given a pattern matrix, which in this case boils down to a binary relation $\Obj \times \Atr \tto{A} 2$,  we consider the induced map $\Atr \tto{\neg A} 2^\Obj$, and derive the monotone maps
\bear
B(X) & = &\{i\in \Obj\ |\ \ \exists u\in X.\ \neg uAi\}\\
B^\ddag (Y) & = &\{u\in \Atr\ |\ \ \forall i\not\in Y.\ uAi\}
\eear
which are adjoint to each other in the sense
\bear
B(X) \subseteq Y & \iff & X\subseteq B^\ddag(Y)
\eear
and by conjugating yield the Galois connection
\bear
Y\subseteq  \neg B(X) & \iff & X\subseteq B^\ddag(\tilde{\neg} Y)
\eear
\[\xymatrix{
2^\Atr  \ar@/^/[rr]^{B}  &&2^\Obj \ar@/^/[rr]^{\neg} \ar@/^/[ll]^{B^\ddag}&& \tilde{2}^\Obj  \ar@/^/[ll]^{\tilde{\neg}}
}\]
The spectral decomposition 
\[\xymatrix{
2^\Atr \ar@/^/[rrrr]^{\neg B} \ar@/_/
@{->>}[dr]_{\bar{V}^\ddag}  && && \tilde{2}^\Obj \ar@/^/[llll]^{B^\ddag \neg}
\ar@/_/@{->>}[dl]_{\bar{W}^\ddag} \\
 & \bar{\AAtr} \ar@{<->}[rr]|{\LLL} \ar@/_/
 @{^{(}->}[ul]_{\bar{V}} & & \bar{\OObj} \ar@/_/
 @{^{(}->}[ur]_{\bar{W}}
}\]
is obtained by setting
\bear
\bar{\AAtr} & =&\{ X \in 2^\Atr \ |\  \Ww^\Atr(X) = X\}\\
\bar{\OObj} & = & \{ Y \in \tilde{2}^\Obj \ |\  \Ww^\Obj(Y) = Y\}
\eear
where the closure operators $\Ww^\Atr =  (B^\ddag\tilde{\neg})\tilde{\circ}(\neg B)$ and $\Ww^\Obj  =   (\neg B)\circ (B^\ddag \tilde{\neg}) $ unfold to 
\bear
\Ww^\Atr(X) & = & \{u\in \Atr\ |\ \forall i\in \Obj.\ (\forall v\in X.\ iAv)\Rightarrow iAu\}\\
\Ww^\Obj(Y) & = & \{i\in \Obj\ |\ \forall u\in \Atr.\ (\forall j\in Y.\ jAu)\Rightarrow iAu\}
\eear
Note that $\Ww^\Atr$ is obtained by composing the matrices $\neg B$ and $B^\ddag\tilde{\neg}$ over the space $\overline{2}^\Obj$, where the composition $\tilde{\circ}$ is dual to the usual one, i.e. $(P\tilde{\circ}Q)_{ik} =  \bigwedge_{j} (P_{ij} \vee Q_{kl})$.
 
It is easy to see that the lattices of closed sets $\bar\AAtr$ and $\bar\OObj$ are isomorphic, because they are both isomorphic with 
\begin{align*}
\LLL =  \big\{<X,Y>\in \WP\Atr\times \WP \Obj\ | & B(X) = \neg Y\ \wedge  \\  &   B^\ddag(Y) = \neg X\big\}
\end{align*}
This is the form in which a {\em concept lattice\/} is usually presented \cite{FCA}. The fact that the spectral composition is {\em minimal\/} means that it correlates users' {\em strongest tastes}, captured in $\bar \AAtr$ with items' {\em strongest styles}, captured in $\bar \OObj$.

\paragraph{Remark.} While LSI is a standard, well-studied data mining method, FCA has been less familiar in the data analysis communities, although an early proposal of a concept-lattice approach can be traced back to the earliest days of the information retrieval research \cite{Salton68}, predating both FCA and even the standard vector space model. More recently, though, the applications of FCA in information retrieval have been tested and explained \cite{Carpineto-Romano,Priss,Marcus07}. The succinct presentation of LSI and FCA as special cases of the same pattern, in our abstract model above, points to the fact that the Singular Value Decomposition, on which LSI is based, and the Galois Connections, that lead to FCA, both subsume under the abstract structure of isometric decomposition, just instantiated to the rig of reals for LSI, and to the boolean rig for FCA. The simple structure of isometric decomposition, and the corresponding notion of conceptual distance, can thus be construed as the basic building block of semantical classification in data analysis.  It turns out that already this rudimentary structure leads into quantum statistics. 

\subsection{Concept lattices are not distributive}
While classical measures are defined over $\sigma$-algebras, which are distributive (and boolean) as lattices, quantum measures are defined over a more general family of algebras, which need not be distributive lattices, but only orthomodular \cite{Meyer86EPQ,Meyer93QProb,Redei06QPT}. 

A crucial, frequently made observation, eventually leading into quantum statistics, is that the lattices of concepts, and of topics, induced by the various forms of latent semantics, are {\em not distributive}. Indeed, since the lattice structure is induced by
\bear
x\wedge y & = & x\cap y\\
x\vee y & = & \Ww(x\cup y)
\eear
the closure operator $M$ often disturbs the distributivity of the underlying set-theoretic operations. The observation that this non-distributivity of concept lattices lifts to the realm of information retrieval is due to van Rijsbergen. For reader's convenience, we repeat the intuitive example of $x \wedge (y\vee z) \neq (x\wedge y) \vee (x \wedge z)$  from \cite[p.~36]{vanRijsbergenGIR}. In a taxonomy of animals,  take $x = $"bird", $y =$ "human" and $z = $"lizzard". Then both $x\wedge y$ and $x\wedge z$ are empty, so that $(x\wedge y) \vee (x \wedge z)$ remains empty. On the other hand, $y\vee z$ = "vertebrates", because vertebrates are the smallest class including both humans and lizzards. Hence $x\wedge (y\vee z)$ = "birds" is not empty.

The point is that such phenomena arise from all forms of latent semantics. But beyond this point, there are even more specific indications of quantum statistics at work.

 \section{Similarity and ranking}
At the core of the vector space model of information retrieval, data mining and other forms of data analysis lies the idea that the basic similarity measure, applicable to pairs of objects, or of attributes, or to the mixtures thereof, is expressible in terms of the inner product of their normalized (often also balanced) vectors:
\bear
\Sim(i,j) & = & <\nor{A}_{j\bullet} | \nor{A}_{i\bullet} >\ = \ \sum_{u\in \Atr} \nor{A}_{ju} \cdot \nor{A}_{iu}\\
\Sim(u,v) & = & <\nor{A}_{\bullet u} | \nor{A}_{\bullet v} >\ = \ \sum_{i\in \Obj} \nor{A}_{iu} \cdot \nor{A}_{iv}
\eear
More generally, using the inner product one can also measure the similarity of pure topics $x$ and $y$, viewed as linear combinations of the property vectors:
\bear
\Sim_M(x,y) & = & <x| A^\ddag A | y> = <Ax | Ay>
\eear
In the same vein, the ranking of mixed topics, represented by the subspaces $E$ of the space of properties, then corresponds to the trace operator:
\bear
\Tr_M (x) & = & <x | A^\ddag A | x> \ = \ <Ax | Ax>\\
\Tr_M (E) & = & \sum_{x\in B_E} \Tr_M (x)
\eear
Noting that a correlation matrix $M = A^\ddag A$ amounts to what is in quantum statistics called an  {\em observable}, we see that the ranking measures, already in the standard vector model, correspond to quantum measures. If the pattern matrices are furthermore normalized as to generate the correlation matrices with a unit trace, then they correspond to quantum probability distributions, or to quantum states.

\section{Bell's inequality of similarities}
In this final section, we attempt to use the described measure of similarity of users' tastes, derived from their past ratings of similar items, to predict the probability that they will agree in their future ratings. Although based on a simple, intuitive view of similarity and agreement, this prediction turns out to be impossible, as it leads to a contradiction. This impossibility result can be viewed as an indicator of a quantum statistical correlation, or at least as evidence that there is a problem with the straightforward statistical model of this simple situation.

The contradiction arises along the lines of Bell's derivation of his notable inequality \cite{Bell64EPR}. More precisely, for any pair of users $x,y\in \Atr$, represented by the unit vectors $x,y : \Obj \to \RRr$, derive from their past ratings of the same items, we consider the random variables $X, Y : \Obj'\to \{0,1\}$, over a possibly larger set of items. Suppose that $X(i) = 1$ means that the user $x$ likes the item $i$, and that $X(i) = 0$ means that she does not like it. We assume that the probability $\Prob(X=Y)\in [0,1]$ that $X$ and $Y$ will agree is proportional to their past similarity $\Sim(x,y)\in [-1,1]$, modulo the rescaling of $[-1,1]$ to $[0,1]$. This induces a constraint on the similarities.

\paragraph{Proposition.} {\em Let the past preferences of $x_0,x_1,y_0,y_1\in \Obj$ be given as unit vectors $x_0,x_1,y_0,y_1 : \Atr \to \RRr$. If the probability of their future agreement is determined by rescaling the similarity of their past preferences 
\bear
\Prob(X=Y) & = & \frac{1+\Sim(x,y)}{2}
\eear
then their similarities must satisfy the following condition:}
\beq\label{Sim}
\Sim(x_0,y_1)+\Sim(x_1,y_1)+\Sim(x_1,y_0) - \Sim(x_0,y_0) \leq  2
\eeq
This follows from the general fact that the disagreement of $\{0,1\}$-valued random variables is a distance function.
\paragraph{Lemma.} {\em Any three random variables $X,Y,Z: \Obj \to \{0,1\}$ satisfy}
\bea\label{Prob}
\Prob(X  \neq Z) & \leq &  \Prob(X\neq Y)+\Prob(Y\neq Z) 
 \eea
 
 \bpr
 Let $W_{XY} : \Atr \to \{0,1\}$ be the random variable
\bear
W_{XY}(i) & = & \begin{cases} 1 & \mbox{ if } X(i) \neq Y(i)\\
0 &  \mbox{ if } X(i) = Y(i)
\end{cases}
\eear
We claim that 
\bea\label{W}
W_{XZ} & \leq & W_{XY} +W_{YZ}
\eea
Towards the contradiction, suppose that there is $j\in \Obj$ with 
$W_{XZ}(j)   \gt  W_{XY}(j) + W_{YZ}(j) $.
This means that $W_{XZ}(j)  =  1$, but $W_{XY}(j) = W_{YZ}(j) = 0$, and thus 
$X(j) \neq Z(j)$ but $X(j) = Y(j)$ and $Y(j) = Z(j)$ ---  which is clearly impossible. Therefore (\ref{W}) must be true. But since $\Prob(X\neq Y) =  \Exp(W_{XY})$, averaging (\ref{W}) gives (\ref{Prob}).
 \epr

\bprf{ of the Proposition} 
Since $\Prob(X=Y) = \frac{1+\Sim(x,y)}{2}$, it follows that $\Prob(X\neq Y) = \frac{1-\Sim(x,y)}{2}$. Substituting this into (\ref{Prob}) gives (\ref{Sim}).
\epr

\paragraph{Corollary.}{\em The probability of users' future agreement $\Prob(X=Y)$ cannot be derived by rescaling the past similarities of their tastes $\Sim(x,y)$, where the similarity measure $\Sim$ is defined by the inner product. The reason is that formula (\ref{Sim}), which would have to be satisfied, does not always hold.}
\bpr
The taste vectors $x_0 = (1,0)$, $y_0 = (-1,0)$, $x_1 = \left(-\frac{1}{2},\frac{\sqrt{3}}{2}\right)$ and $y_1 = \left(\frac{1}{2},\frac{\sqrt{3}}{2}\right)$ provide a counterexample for (\ref{Sim}).
\epr

\paragraph{Interpretation.} Why is it not justified to predict future agreements from past similarities, both defined in intuitively obvious ways?
One line of explanation is that {\em the independence assumptions are violated}. As usually, the dependencies can be explained in terms of hidden variables (e.g., off-line interactions of the users), or in terms of non-local interactions.
Another line of explanation is that {\em the dependencies are introduced in the model itself}. Intuitively, this means that the users, whose agreements are predicted, have not been sampled in the same measure space, and that their preferences should not be statistically mixed.

\paragraph{Remark.} 
Rather than derived from similarity, users' semantical distance can be defined by $\Prob(X=Y)  =  |Ax - Ay|_\infty$. A reader familiar with quantum probability theory \cite{Meyer86EPQ,Meyer93QProb} will recognize this interaction of the Hilbert space $\ell_2$ and the Banach space $\ell_\infty$, which acts on it as a von Neumann algebra, as the familiar interface between the quantum and the classical probabilities.

\section{Conclusion and future work} 
We have shown that already in the basic, but sufficiently abstract models of information retrieval, data mining, and other forms of data analysis, a suitable version of Bell's argument applies, suggesting that the quantum statistical approach may be necessary. 

The simple interpretation of Bell's argument is that the quantum statistical predictions refer to non-local interactions. More subtle interpretations lead into the issues of contextuality \cite[p.~9]{Bell:book}. In some cases, of course, both the non-local interactions and the contextual dependencies arise as a figment of the statistical model, mixing variables that cannot be sampled together. Either way, the version of the argument presented above suggests simple minded prediction based on  the vector space model of information processing in a network may lead to problems if the locality of the interactions is not taken into account. Is it possible that genuine entanglement phenomena arise on a network?

After a moment of thought about this question, one gets a strange feeling that quantum probability might in fact be easier to comprehend in the realm of network computation, than in physics.\footnote{Perhaps like the theory of parallel universes, which seems to have more convincing interpretations in everyday life, and in distributed computation, than in physics.} While action at a distance is a highly unintuitive phenomenon in physics --- Einstein called it "spooky" --- in network computation it can be reduced to the fact that the information may flow not only through the network links, but also off the network. This fact is not only intuitively natural, in the sense that, say, the data on the Web move not only in packets, along the Internet links, but they also get teleported from site to site, by people talking to each other, and then typing on their keyboards; but it is also information theoretically robust, in the sense that there are always covert channels. In abstract models, they can be represented in terms of non-local hidden variables, or in terms of entanglement. Either way, the operational content of quantum statistical methods will undoubtedly broaden the algorithmic horizons of network computation and data analysis, already by analyzing the meaning of the notable quantum algorithms in physics-free implementations. Convenient toolkits for combining quantum states, and for composing quantum operations \cite{PavlovicD:QMWS} are likely to acquire new roles in latent semantics. On the other hand, the generic no-cloning and no-broadcasting theorems \cite{barnum06cloning} are likely to point to some interesting statistical limitations, with a potential impact in security.\footnote{One direct consequence of the no-cloning theorem seems to be that only classical styles can be copied.} 

\paragraph{Acknowledgement.} I am grateful to Eleanor Rieffel for pointing out an error in an earlier version of this abstract, caused by some of my notational abuses.

\bibliography{ref-qsearch}

\begin{thebibliography}{}

\bibitem[\protect\citeauthoryear{Abramsky \& Coecke}{2004}]{Abramsky-Coecke}
Abramsky, S., and Coecke, B.
\newblock 2004.
\newblock A categorical semantics of quantum protocols.
\newblock In {\em Proceedings of the 19th Annual IEEE Symposium on Logic in
  Computer Science (LICS)}.
\newblock IEEE Computer Society.
\newblock Also arXiv:quant-ph/0402130.

\bibitem[\protect\citeauthoryear{Azar \bgroup \em et al.\egroup
  }{2001}]{azar01spectral}
Azar, Y.; Fiat, A.; Karlin, A.~R.; McSherry, F.; and Saia, J.
\newblock 2001.
\newblock Spectral analysis of data.
\newblock In {\em {ACM} Symposium on Theory of Computing},  619--626.

\bibitem[\protect\citeauthoryear{Barnum \bgroup \em et al.\egroup
  }{2006}]{barnum06cloning}
Barnum, H.; Barrett, J.; Leifer, M.; and Wilce, A.
\newblock 2006.
\newblock Cloning and broadcasting in generic probabilistic theories.

\bibitem[\protect\citeauthoryear{Bell}{1964}]{Bell64EPR}
Bell, J.~S.
\newblock 1964.
\newblock On the {Einstein-Podolsky-Rosen} paradox.
\newblock {\em Physics} 1:195--200.

\bibitem[\protect\citeauthoryear{Bell}{1987}]{Bell:book}
Bell, J.~S.
\newblock 1987.
\newblock {\em Speakable and Unspeakable in Quantum Mechanics}.
\newblock Cambridge University Press.

\bibitem[\protect\citeauthoryear{Carpineto \& Romano}{2004}]{Carpineto-Romano}
Carpineto, C., and Romano, G.
\newblock 2004.
\newblock Exploiting the potential of concept lattices for information
  retrieval with credo.
\newblock {\em Journal of Universal Computer Science} 10(8):985--1013.

\bibitem[\protect\citeauthoryear{Coecke \& Pavlovic}{2007}]{PavlovicD:QMWS}
Coecke, B., and Pavlovic, D.
\newblock 2007.
\newblock Quantum measurements without sums.
\newblock In Chen, G.; Kauffman, L.; and Lamonaco, S., eds., {\em Mathematics
  of Quantum Computing and Technology}. Taylor and Francis.
\newblock  559--596.

\bibitem[\protect\citeauthoryear{Deerwester \bgroup \em et al.\egroup
  }{1990}]{Deerwester90indexing}
Deerwester, S.~C.; Dumais, S.~T.; Landauer, T.~K.; Furnas, G.~W.; and Harshman,
  R.~A.
\newblock 1990.
\newblock Indexing by latent semantic analysis.
\newblock {\em Journal of the American Society of Information Science}
  41(6):391--407.

\bibitem[\protect\citeauthoryear{Everitt}{1984}]{Everitt84latent}
Everitt, B.
\newblock 1984.
\newblock {\em An Introduction to Latent Variable Models}.
\newblock London: Chapman \& Hall.

\bibitem[\protect\citeauthoryear{Ganter, Stumme, \& Wille}{2005}]{FCA}
Ganter, B.; Stumme, G.; and Wille, R., eds.
\newblock 2005.
\newblock {\em Formal Concept Analysis, Foundations and Applications}, volume
  3626 of {\em Lecture Notes in Computer Science}. Springer.

\bibitem[\protect\citeauthoryear{Jolliffe}{1986}]{Jolliffe86PCA}
Jolliffe, I.~T.
\newblock 1986.
\newblock {\em Principal Component Analysis}.
\newblock Springer Series in Statistics. Springer-Verlag.

\bibitem[\protect\citeauthoryear{Kleinberg}{1999}]{kleinberg99authoritative}
Kleinberg, J.~M.
\newblock 1999.
\newblock Authoritative sources in a hyperlinked environment.
\newblock {\em Journal of the ACM} 46(5):604--632.

\bibitem[\protect\citeauthoryear{Meyer}{1986}]{Meyer86EPQ}
Meyer, P.-A.
\newblock 1986.
\newblock \'{E}l\'{e}ments de probabilit\'{e}s quantiques (expos\'{e}s {I}
  \`{a} {IV}).
\newblock In {\em S\'{e}minaire de probabilit\'{e}s de Strassbourg}, volume
  1204,1247 of {\em Lecture Notes in Mathematics}. Berlin: Springer-Verlag.

\bibitem[\protect\citeauthoryear{Meyer}{1993}]{Meyer93QProb}
Meyer, P.-A.
\newblock 1993.
\newblock {\em Quantum Probability for Probabilists}.
\newblock Number 1538 in Lecture Notes in Mathematics. Springer-Verlag.

\bibitem[\protect\citeauthoryear{Page \bgroup \em et al.\egroup
  }{1998}]{page98pagerank}
Page, L.; Brin, S.; Motwani, R.; and Winograd, T.
\newblock 1998.
\newblock The {PageRank} citation ranking: Bringing order to the {Web}.
\newblock Technical report, Stanford Digital Library Technologies Project.

\bibitem[\protect\citeauthoryear{Poshyvanyk \& Marcus}{2007}]{Marcus07}
Poshyvanyk, D., and Marcus, A.
\newblock 2007.
\newblock Combining formal concept analysis with information retrieval for
  concept location in source code.
\newblock In {\em ICPC '07: Proceedings of the 15th IEEE International
  Conference on Program Comprehension},  37--48.
\newblock Washington, DC, USA: IEEE Computer Society.

\bibitem[\protect\citeauthoryear{Priss}{2006}]{Priss}
Priss, U.
\newblock 2006.
\newblock Formal concept analysis in information science.
\newblock In Cronin, B., ed., {\em Annual Review of Information Science and
  Technology}, volume~40.

\bibitem[\protect\citeauthoryear{Redei \& Summers}{2006}]{Redei06QPT}
Redei, M., and Summers, S.~J.
\newblock 2006.
\newblock Quantum probability theory.
\newblock To appear in {\em Studies in the History and Philosophy of Modern
  Physics}.

\bibitem[\protect\citeauthoryear{Salton}{1968}]{Salton68}
Salton, G.
\newblock 1968.
\newblock {\em Automatic Information Organization and Retrieval.}
\newblock McGraw Hill Text.

\bibitem[\protect\citeauthoryear{{van Rijsbergen}}{2004}]{vanRijsbergenGIR}
{van Rijsbergen}, C.~J.
\newblock 2004.
\newblock {\em The Geometry of Information Retrieval}.
\newblock New York, NY, USA: Cambridge University Press.

\bibitem[\protect\citeauthoryear{Wille}{1982}]{Wille82FCA}
Wille, R.
\newblock 1982.
\newblock Restructuring lattice theory: an approach based on hierarchies of
  concepts.
\newblock In Rival, I., ed., {\em Ordered Sets}. Dordrecht: Dan Reidel.
\newblock  445--470.

\end{thebibliography}
\bibliographystyle{aaai}
\end{document}